\documentstyle[aps,epsfig]{revtex}                         % submission          %
\draft
\begin{document}
\twocolumn[\hsize\textwidth\columnwidth\hsize
     \csname @twocolumnfalse\endcsname
%\oddsidemargin -0.35in
%\evensidemargin -0.35in

%\preprint{}%

%------------------------------------------------------------------------------%
%       T I T L E                                                              %
%------------------------------------------------------------------------------%

\title{Visibility, efficiency, and Bell violations in real
       Einstein-Podolsky-Rosen experiments.
       }

\author{W. A. Hofer}
\address{
         Department of Physics and Astronomy,
         University College, Gower Street, London WC1E 6BT, UK}

\maketitle

\begin{abstract}
The violation of Bell's inequalities in Einstein-Podolsky-Rosen
experiments has been demonstrated for photons and ions. In all
experiments of this kind the relation between visibility,
efficiency, and Bell violation is generally unknown. In this paper
we show that simulations based on a local hidden variables models
for entangled photons provide this information. It is established
that these properties are closely related by the way, in which
photons are detected after a polarizer beam splitter. On this
basis we suggest controlled experiments which, for the first time,
subject the superposition principle to experimental tests.
\end{abstract}


\vskip2pc]

The Einstein-Podolsky-Rosen (EPR) problem has long occupied a
central place in the understanding of quantum mechanics
\cite{epr35,bell64,aspect82,weihs98,aspect99,rowe01}. Bell's
inequalities in conjunction with correlation measurements seemed
to prove that reality in microphysics is manifestly nonlocal
\cite{bell64,aspect82,weihs98,aspect99}. Furthermore, the
experimental evidence seems to contradict even the notion of an
independent reality \cite{espagnat89}.

But as recently shown, the experimental data can be fully
reproduced with a local and realistic model of correlation
measurements \cite{hofer01a}. The model focussed on the
electromagnetic fields of the propagating photons. The essential
connection between the two points of measurement is the phase
acquired at their common origin. The basic realistic model,
Furry's integral \cite{furry36}, was shown to be a wrong
representation of the digital output of the polarizer beam
splitters (PBS). The new model has three key advantages: (i) It is
a hidden variables model, the hidden variable is the angle of
polarization of a photon's electromagnetic field; (ii) it is
strictly local, all events can be traced from one point in space
and time to the next; (iii) the main experimental limitations are
included in the model. Decoherence, for example, is described as a
random segment of a photon's optical path. We also introduced a
generic parameter, the threshold $\Delta s$, which describes the
data selection at the critical polarization angle 45$^{\circ}$ of
a PBS. In simulations of actual experiments it was shown that the
Bell inequalities can be violated by a close to arbitrary amount,
depending on the data suppressed around the critical angle.

This result points to a gap in our current understanding of these
important experiments. The relation between visibility,
decoherence, Bell violations, and data selection is in general not
known. The problem is aggravated by the aim of experimenters for
high visibility in their experiments. Because this, in turn, may
destroy the fair sampling of their experimental data. The most
efficient method to gain a proper understanding of the relation
between the important parameters in such an experiment are
numerical simulations.

In this paper we simulate experiments over the whole range of
experimental parameters. The parameter space is determined by
decoherence and by the value of $\Delta s$. We simulate
measurements from fully coherent to fully decoherent photon beams.
The PBS parameters in one limit ($\Delta s = 0$) allow an
arbitrarily precise resolution of angles at 45$^{\circ}$. In the
opposite limit ($\Delta s = 0.5$) not a single photon will be
measured, because the limit is outside the range of possible
values. The paper is structured as follows: first we shall briefly
describe the setup of the experiment and explain, why the diagonal
in a PBS is such a critical angle. Then we shall introduce the
numerical model and explain its connection to experimental
parameters. And finally, we shall present our simulations of EPR
experiments under realistic conditions over the whole parameter
space of the model.

The main idea which underlies the numerical model is a one to one
correspondence between the model and the experiment for {\em
single} events. It is, for example, conceivable that agreement
between experiments and simulations could be obtained only after
all results have been summed up. In this case the statistics enter
the picture in the distribution of single events, which may be
different for the model and the experiment. In fact, no successful
model on this basis has ever been developed. On the contrary, the
only local realistic model within this framework \cite{furry36}
does not reproduce experiments. In this model the total
probability $P(\alpha,\beta)$ of a coincidence between two
polarizers set to $\alpha$ and $\beta$, respectively, is given by
the sum over $\cos^2(\lambda - \alpha) \cos^2(\lambda - \beta)$,
where $\lambda$ is the hidden angle of polarization. The reason
for the disagreement is that in the experiments we do not sum up
products of trigonometric functions, which in principle can be
arbitrarily low, but we sum up integer coincidences on both
measurement devices. In effect, we make the results digital on the
level of single photons and thus introduce a cutoff in our
trigonometric functions. This cutoff is due to the PBS.

In experiments with entangled photons a pair of photons is emitted
from a common source. After the polarization of one photon is
altered by an angle $\alpha$ both photons are analyzed at their
respective PBS. In our model we simulate this feature by a switch
at the critical angle of the PBS of $|\lambda - \alpha| =
45^{\circ}$. The polarizer beam splitter projects a given field
vector of the electromagnetic field onto two orthogonal
directions, say $\alpha_{1}$ and $\alpha_{2}$. The output of one
channel is the projection onto one axis (+), the intensity is
$\cos^2(\phi_{1} - \alpha_{1})$, where $\phi_{1}$ is the
polarization angle of the electromagnetic field. The projected
intensity onto the other axis (-) is $\sin^2(\phi_{1} -
\alpha_{1})$, since $\alpha_{2}$ is perpendicular to $\alpha_{1}$.
If we set $\alpha_{1} = 0$ and $\alpha_{2} = \pi/2$, then if
$\phi_{1} = \pi/4$ we get the same intensity on both outputs (+)
and (-). This case must be excluded, since it is not compatible
with the desired results in quantum mechanics, where we have
either (+) or (-), but never both. For this reason we include a
threshold to account for the elimination of undesired results. The
threshold $\Delta s$ simulates the way, the combination of a PBS
and a detector operates. Statistics enter our model due to an
unknown initial polarization of the coupled two-photon system. The
unknown initial phase is created by a random number generator
\cite{nr87}. It is also the hidden variable, which makes the
single event unpredictable.

It should be noted that the probability $P(\alpha,\beta)$ of the
ideal measurement can also be described as an integral of two
factored functions $\bar{A}(\phi_1,\alpha)$ and
$\bar{B}(\phi_2,\alpha)$. In the ideal case these functions for
photons of perpendicular polarization ($\phi_2 = \phi_1 + \pi/2$)
are given by:

\begin{equation}
\bar{A}(\phi_1,\alpha) = \frac{1}{2} \left[ sign \left[
\cos^2(\phi_1 - \alpha) - \frac{1}{2}\right] + 1 \right]
\end{equation}
\begin{equation}
\bar{B}(\phi_2,\beta) = \frac{1}{2} \left[ sign \left[
\sin^2(\phi_1 - \beta) - \frac{1}{2}\right] + 1 \right]
\end{equation}
\begin{equation}
P(\alpha,\beta) = \frac{1}{2 \pi} \int_0^{2 \pi} d \phi_1
\bar{A}(\phi_1,\alpha) \bar{B}(\phi_2,\beta)
\end{equation}

The model is deterministic in the sense that it is uniquely
determined by the angles $\alpha$ and $\beta$, and that it is a
unique integral over the angle of polarization $\phi_{1}$ (the
local hidden variable). In this formulation we encounter no longer
any element of randomness. But the model is also compatible with
the fundamental assumption in Bell's inequalities. The
inequalities, however, are not compatible with quantum mechanics,
because they entail infinitely precise measurements even for
non-commuting observables. Since operators of different settings
do not commute, the spin states at different settings cannot be
simultaneously eigenstates of the system. Thus it is impossible to
obtain the limit of the standard inequalities within the framework
of quantum mechanics.

A simulation run starts by an initialization of the random number
generator \cite{nr87}. The generator is initialized only once, at
the beginning of a simulation cycle. The random number is mapped
onto the initial phase from 0 to $2 \pi$ of the photon pair.
Simulations are generally made with a phase difference of $\pi/2$
between the angles of polarization of photon one and photon two.
We introduce decoherence as a random segment of the optical path
of a photon. 100 \% decoherence, for example, means that half a
wavelength of the optical path is random. After covering the
distance to polarizer one and two the photons are measured. We
assumed, without lack of generality, that both distances are
integer multiples of the wavelength. After a single pair has been
measured, we record the coincidences \mbox{(++,+-,-+,--)}. The
procedure is repeated for all pairs, then the polarization angle
of device one is changed by $\pi/100$. A run ends, when all pairs
at the final position of polarizer one have been measured ($\pi$).
In the simulation we recorded the correlations of 2000 pairs of
photons. We vary decoherence from 0\% to 100\%, and the threshold
of the polarizer from 0.0 to 0.5. The results displayed, the
correlation function, the visibility, and efficiency plots are all
derived from $N_{++}$.

\begin{figure}
\begin{center}
\epsfxsize=0.9\hsize \epsfbox{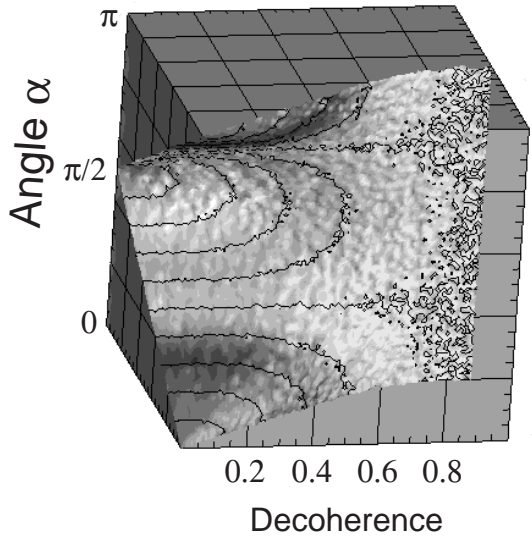}
\end{center}
\vspace{0.5 cm} \caption{Variation of the correlation function
         with decoherence. The value of the correlation function
         is color coded. For fully coherent field we obtain a
         sawtooth, while fully decoherent photon beams will show
         only statistical noise.
        }
\label{fig001}
\end{figure}

One key requirement in EPR experiments is the space like
separation of the two measurements. In general this involves an
optical waveguide between the source of the photon pair and the
polarizers. Unless all components in such an experiment are cooled
to temperatures near 0K and decoupled from their environment, we
expect the components of the optical system to oscillate. This
oscillation introduces randomness into the correlation
measurements, which has the same effect as decoherence of the
electromagnetic field of the photon. Any realistic theoretical
model must include a random optical path of at least 5\% or about
10 - 20nm, if the laser operates with visible light. This
translates into about 10\% decoherence. Depending on the setup and
the experimental precision much higher decoherence seems possible.
In Fig. \ref{fig001} we show the correlation function for the
whole range of decoherence from 0\% (fully coherent fields) to
100\% (fully random polarization angles). As expected, the
correlation function becomes a straight line in the limit of full
decoherence. In the intermediate region (10 - 50 \% decoherence)
it is sinusoidal, while the ideal correlation is a sawtooth
\cite{hofer01a}. It should be noted that the phase is a hidden
variable in all measurements, also of linear polarization. Because
the polarization measurements at a device which combines polarizer
beam splitters and photodetectors is an energy measuring device.
And energy, from a field theoretical point of view, is intensity,
which depends on the phase at the point of measurement.

\begin{figure}
\begin{center}
\epsfxsize=0.9\hsize \epsfbox{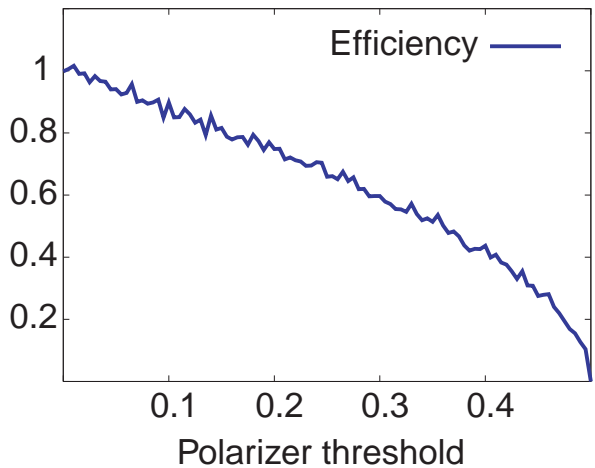}
\end{center}
\vspace{0.5 cm} \caption{Efficiency of an EPR experiment for
        varying polarizer threshold. The efficiency decreases
        as the threshold is increased.
        }
\label{fig002}
\end{figure}

In all experiments efficiency is less than 10\% \cite{weihs98}.
Within the present model efficiency decreases due to increased
polarizer thresholds. This decrease is close to linear in the high
efficiency range. Efficiency in terms of the polarizer threshold
is plotted in Fig. \ref{fig002}. The curve shows the statistical
noise due to the random variables in our simulation. Given the
values we obtain, the main reason for the lack of efficiency in
the measurements should not be the polarizer threshold. For
realistic values around 0.1, values which can be deduced from the
visibility in a measurement\cite{hofer01a}, the efficiency is
still around 80\%. We thus have to conclude that the experimental
settings are not the main reason for lacking efficiency. However,
if the efficiency decreases with a change of experimental
parameters, one possible reason is the increase of the polarizer
threshold. And this, in turn, changes the most important values
measured in such an experiment.

\begin{figure}
\begin{center}
\epsfxsize=0.9\hsize \epsfbox{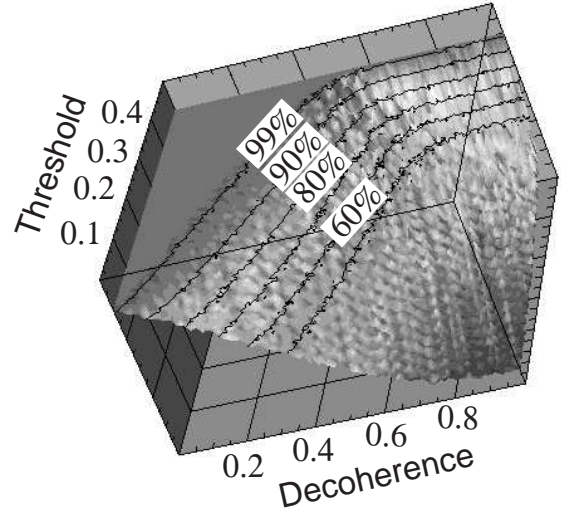}
\end{center}
\vspace{0.5 cm} \caption{Visibility in an EPR experiment. Even
        in experiments, where the photon beams are not fully
        coherent, we can obtain close to 100 \% visibility by
        increasing the polarizer threshold.
        The visibility is color coded.
        }
\label{fig003}
\end{figure}

One of these values is the visibility of the correlation function.
It is defined as $(max - min)/(max + min)$, where $max$ ($min$)
denote the maximum (minimum) number of coincidences over all
polarizer settings. Essentially, the visibility can be increased
by reducing the minimum count to a value close to zero. The
increase of the polarizer threshold has exactly this effect. In
Fig. \ref{fig003} we show the result of our simulation. The
 visibility is color coded and emphasized by
contour lines. Even in case of decoherence of more than 20\% it is
possible to obtain 95\% visibility in the experiment if the
polarizer threshold is raised to 0.13. The experimental values of
Weihs et al. \cite{weihs98}, for example, point to a decoherence
of 10 -- 12\% and a threshold of 0.1 -- 0.12. It seems quite
unexpected that the parameter range with a visibility of more than
99\% covers roughly one fourth of the whole parameter space. The
requirement of high visibility therefore does not decisively limit
the parameter space of real measurements. Visibility seems thus
unsuitable as a measure of experimental precision. As a technical
point we remark that the statistical spread from one value to the
next shows a high fluctuation. This is due to a multiplication of
statistical deviations in the computation of the visibility.

\begin{figure}
\begin{center}
\epsfxsize=0.9\hsize \epsfbox{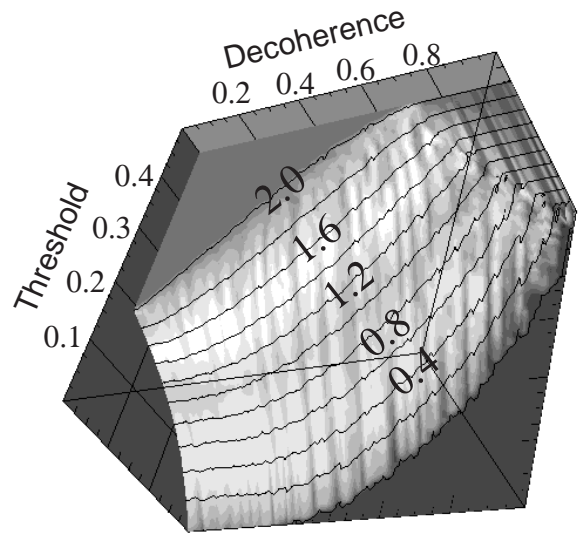}
\end{center}
\vspace{0.5 cm} \caption{Violation of the Bell inequalities
           depending on decoherence and polarizer threshold.
           The violation is color coded.
           The Bell inequalities are violated in all realistic
           experimental setups. The violation depends crucially
           on the experimental parameters and can reach a value
           of up to 2.0.
        }
\label{fig004}
\end{figure}

Experimentally, one seeks to obtain high visibility in an
experiment before embarking on the actual run, where valid data
are measured. In this case it is to be expected that practically
all the data in these experiments are obtained in the high
visibility range. Incidentally, this is the same range where the
Bell inequalities are maximally violated. For the calculation of
the Bell violations we fixed polarizer one and two at the angles
of maximum violation
(0$^{\circ}$,22.5$^{\circ}$,45$^{\circ}$,67.5$^{\circ}$), and
simulated the measurement of 10000 pairs of photons for every
datapoint. We computed the Bell violation from the coincidence
counts using the formulation of Clauser et al. \cite{clauser69}.
From this value we have subtracted 2.0 and show only the values
above zero. The simulation is shown in Fig. \ref{fig004}. It can
be seen that every threshold allows for a wide range of Bell
violations. For a threshold value $> 0.2$, for example, the
inequalities can be violated by 0.0 -- 2.0, depending on the
decoherence of the photon beams. The contour lines describe the
violation. Comparing with the previous plot it can be seen that
the condition of high visibility alone guarantees that the Bell
inequalities are maximally violated. If a violation exceeding 0.82
(the value in quantum mechanics) has so far not been observed, we
refer this to the low coherence in the measurements and the
requirement of high efficiency. The former generally reduces the
violation, while the latter makes a high threshold undesirable.

A synopsis of all data obtained in the simulations reveals that
the experimental results can be changed arbitrarily by changing
the parameters in the experiments. Generally speaking, EPR
experiments if analyzed with our model exhibit a much larger span
of possible parameters and results than presently assumed. From
our simulations we conclude that there will be not one value for
the Bell violation, which can be compared with theoretical
predictions, but there should be many, depending entirely on the
parameters in the experiment. In this sense the simulation allows,
for the first time, to check quantum mechanics in controlled EPR
experiments. If the Bell violation decreases from an initial value
 as experimental conditions become more ideal, then the predictions
 in quantum mechanics must be wrong. Increasing coherence can
be obtained e.g. by cooling down the experimental components to
very low temperature and by reducing their distances to a minimum.
If, however, increasing coherence and efficiency leads to a higher
violation of the Bell inequalities, then the present local model
of EPR measurements must be inadequate. The actual point in the
parameter space of such an experiment can be obtained from
experimental values. Increasing visibility and efficiency is
related to to lower threshold values and lower decoherence, in
short, to more ideal experimental conditions. With these
experiments also the superposition principle can be tested
experimentally. The principle is fundamental for all current
standard frameworks, it is based on linearity of fields and
wavefunctions between two points of measurement. The principle is
violated in our model, because the two measurements have been
factored. We can therefore use the predictions of our theoretical
model for a first experimental test of superposition. In case the
local hidden variables model is correct, which can be checked by
the above procedure, the superposition principle is not generally
applicable: a consequence which may require major adjustment in
current theories.

In summary we have shown that the whole parameter space of EPR
experiments can be analyzed  by numerical simulations within a
local hidden variables model. We pointed out that the Bell
inequalities are maximally violated in all cases, where an
experiment possesses high visibility. And we suggested controlled
EPR experiments to detect a possible deviation of quantum
mechanical predictions from the results of measurements.

{\em Acknowledgements.} - The EPR problem has been widely
discussed with colleagues over the last year. In particular I am
indebted to G. Adenier, D. Bowler, A. Fisher, J. Gavartin, J.
Gittings, and R. Stadler. Computing facilities at the HiPerSPACE
center were funded by the Higher Education Funding Council for
England.

%------------------------------------------------------------------------------%
%      R E F E R E N C E S
%
%------------------------------------------------------------------------------%

%------------------------------------------------------------------------------%
%      F I G U R E S
%
%------------------------------------------------------------------------------%

\begin{references}
\bibitem{epr35} A.
Einstein, B. Podolsky, and N. Rosen, {\it Phys. Rev.} {\bf 47},
777 (1935)
\bibitem{bell64}
J.S. Bell, {\it Physics} {\bf 1}, 195 (1964)
\bibitem{aspect82}
A. Aspect, P. Grangier, and G. Roger, {\it Phys. Rev. Lett.} {\bf
61}, 91 (1982)
\bibitem{weihs98}
G. Weihs, T. Jennewein, C. Simon, H. Weinfurter, and A. Zeilinger,
{\it Phys. Rev. Lett.} {\bf 81}, 5039 (1998)
\bibitem{aspect99}
A. Aspect, {\it Nature} {\bf 398}, 189 (1999)
\bibitem{rowe01}
M. A. Rowe et al. {\it Nature} {\bf 409}, 791 (2001)
\bibitem{espagnat89} B.
d'Espagnat, {\it Conceptual Foundations of Quantum Mechanics},
Addison Wesley, New York (1989)
\bibitem{hofer01a}
W.A. Hofer, Simulation of Einstein-Podolsky-Rosen experiments in a
local hidden variables model with limited efficiency and
coherence, {\it Phys. Rev. Lett.}, submitted (2001)
\bibitem{furry36} W.H.
Furry, {\it Phys. Rev.} {\bf 49}, 393 (1936)
\bibitem{nr87}
W.H. Press, B.P.Flannery, S.A. Teukolsy, W.T. Vetterling, {\it
Numerical Recipes}, Cambridge Univ. Press, Cambridge (1987)
\bibitem{clauser69}
J.F. Clauser, M.A. Horne, A. Shimony, and R.A. Holt, {\it Phys.
Rev. Lett.} {\bf 23}, 880 (1969)
\end{references}
\end{document}